\documentclass{article}
\usepackage{amsfonts}
\usepackage{amsmath}

\newtheorem{theorem}{Theorem}
\newtheorem{acknowledgement}[theorem]{Acknowledgement}

\input{tcilatex}

\begin{document}

\title{Equations Relating Structure Functions of all Orders}
\author{Reginald J. Hill \\
National Oceanic and Atmospheric Administration/\\
Environmental Research Laboratory,\\
Boulder CO, USA, 80305-3328}
\maketitle

\begin{abstract}
Exact equations are given that relate velocity structure functions of
arbitrary order with other statistics. \ \textquotedblleft
Exact\textquotedblright\ means that no approximations are used except that
the Navier-Stokes equation and incompressibility condition are assumed to be
accurate. \ The exact equations are used to determine the structure function
equations of all orders for locally homogeneous but anisotropic turbulence
as well as for the locally isotropic case. \ The uses of these equations for
investigating the approach to local homogeneity as well as to local isotropy
and the balance of the equations and identification of scaling ranges are
discussed. \ The implications for scaling exponents and investigation of
intermittency are briefly discussed.
\end{abstract}

\section{\textbf{Introduction}}

\qquad Kolmogorov's (1941) equation and Yaglom's equation were the first two
equations of the ``dynamic theory'' of the local structure of turbulence. \
The name ``dynamic theory'' was originated by Monin \& Yaglom (1975) (their
Sec. 22) to mean the derivation of equations relating structure functions by
use of the Navier-Stokes equation and/or the scalar conservation equation,
and the investigation of the resulting statistical equations. \ Monin \&
Yaglom (1975) pointed out that the dynamic theory gives important
relationships among structure functions, and that these relationships
provide extensions of predictions based on dimensional analysis. \
Theoretical studies (Lindborg 1996; Hill 1997a) clarified the assumptions
that are the basis of Kolmogorov's equation and give equations that are
valid for anisotropic and locally homogeneous turbulence as well as for the
case of local isotropy and local homogeneity. \ Antonia \textit{et al}.
(1983) and Chambers \& Antonia (1984) used experimental data to study of the
balance of the classic equations of Kolmogorov and Yaglom. \ There is
renewed interest in examining the balance of those equations using both
experimental and DNS data and in generalizing the equations to cases of
inhomogeneous, nonstationary, and anisotropic turbulence (Lindborg 1999;
Danaila \textit{et al}., 1999a,b,c; Antonia \textit{et al}. 2000). \ Whereas
Kolmogorov's (1941) equation relates 2nd- and 3rd-order velocity structure
functions, the next-order dynamic equation relates 3rd- and 4th-order
structure functions and a pressure-gradient, velocity-velocity structure
function. \ The balance of that next-order equation has been examined by
means of experimental and DNS data; this showed the behavior of the
pressure-gradient, velocity-velocity structure function (Hill \& Boratav
2001). \ There is now interest in dynamic-theory equations of arbitrarily
high order $N$ (Yakhot 2001). \ Such equations relate velocity structure
functions of order $N$\ and $N+1$\ and other statistics. \ Those equations
are given in this paper.

\qquad Using the assumptions of local homogeneity, local isotropy and the
Navier-Stokes equation, Yakhot (2001) derived the equation for the
characteristic function of the probability distribution of two-point
velocity differences. \ He uses that equation to derive higher-order dynamic
equations. \ Equations for arbitrarily high-order structure functions can be
obtained\ by repeated application of his differentiation procedure. \ Yakhot
(2001) studies the inertial-range, deduces a closure, and thereby determines
the inertial-range scaling exponents of velocity structure functions. \
Yakhot's study is the first to make significant use of dynamic-theory
equations to determine scaling exponents.

\qquad The purposes and theoretical method of the present paper differ from
those of Yakhot (2001), but one purpose is to verify Yakhot's equations from
our distinctly different derivation. \ That verification is given in Sec. 5.
\ In Sec. 2, exact statistical equations relating velocity structure
functions of any order are derived from the Navier-Stokes equation. \
``Exact'' means that no assumptions are made other than the assumption that
the Navier-Stokes equation and incompressibility are accurate. \ Since the
equations are exact, they apply to any flow, including laminar flow and
inhomogeneous and anisotropic turbulent flow. \ The exact statistical
equations can be used to verify DNS computations and detect their
limitations. New experimental methods of Dahm and colleagues (Su \& Dahm
1996) can also be tested. \ For example, if DNS data are used to evaluate
the exact statistical equations, then the equations should balance to within
numerical precision, otherwise a computational problem is indicated. \ In
Sec. 3, statistical equations valid for locally homogeneous and anisotropic
turbulence are obtained from the exact equations; those equations can be
used with DNS or experimental (Su \& Dahm 1996) data to study the approach
to local homogeneity of a particular flow. \ This can be done by quantifying
the terms that are neglected when passing from exact equations to the
locally homogeneous case, and by quantifying changes in the retained terms
as local homogeneity is approached when the spatial separation vector is
decreased. \ In Sec. 4, statistical equations valid for locally isotropic
and locally homogeneous turbulence are obtained from those for the locally
homogeneous case. \ The approach to local isotropy can be studied by means
analogous to the above described evaluation of local homogeneity. \ Such
studies might shed light on the observed persistence of anisotropy (Pumir \&
Shraiman 1995; Shen \& Warhaft 2000). \ All dynamic-theory equations are now
available to extend the above-mentioned previous studies of the balance of
dynamic-theory equations.

\qquad Incompressibility requires that the different components of the
second-order velocity structure function have the same scaling exponent in
the inertial range. \ The same is true for the third-order structure
function. \ However, at fourth and higher order there is no such
requirement. \ There have been many studies of the possibility that the
inertial-range scaling exponents of the various structure-function
components are unequal (e.g., Chen \textit{et al.} 1997; Boratav \& Pelz
1997; Boratav 1997; Grossmann \textit{et al.} 1997; van de Water \&
Herweijer 1999; Camussi \& Benzi 1997; Dhruva \textit{et al.} 1997; Antonia 
\textit{et al.} 1998; Kahaleras \textit{et al.} 1996; Noullez \textit{et al.}
1997; Nelkin 1999; Zhou \& Antonia 2000; Kerr \textit{et al.} 2001). \ The
usefulness of applying the higher-order dynamic-theory equations to those
investigations is considered in Sec. 6.

\qquad Derivation of the equations produces substantial mathematical detail.
\ Matrix-based algorithms are invented such that the isotropic formulas for
the divergence and Laplacian of isotropic tensors of any order can be
generated by computer. \ The details of this mathematics are available and
are herein referred to as the Archive.\footnote{%
The document ``Mathematics of structure-function equations of all orders''
by R. J. Hill is available from the editorial archive of the Journal of
Fluid Mechanics and at xxx.lanl.gov.}

\section{\textbf{Exact two-point equations}}

\qquad The Navier-Stokes equation for velocity component $u_{i}(\mathbf{x}%
,t) $ is 
\begin{equation}
\partial _{t}u_{i}(\mathbf{x},t)+u_{n}(\mathbf{x},t)\partial _{x_{n}}u_{i}(%
\mathbf{x},t)=-\partial _{x_{i}}p(\mathbf{x},t)+\nu \partial
_{x_{n}}\partial _{x_{n}}u_{i}(\mathbf{x},t),  \label{1NSE}
\end{equation}%
and the incompressibility condition is $\partial _{x_{n}}u_{n}(\mathbf{x}%
,t)=0$. \ In (\ref{1NSE}), $p(\mathbf{x},t)$ is the pressure divided by the
density (density is constant), $\nu $ is kinematic viscosity, and $\partial $
denotes partial differentiation with respect to its subscript variable. \
Summation is implied by repeated\ Roman indices. \ Consider another point $%
\mathbf{x}^{\prime }$ such that $\mathbf{x}^{\prime }$\ and $\mathbf{x}$ are
independent variables. \ For brevity, let $u_{i}=u_{i}(\mathbf{x},t)$, $%
u_{i}^{\prime }=u_{i}(\mathbf{x}^{\prime },t)$, etc. Require that $\mathbf{x}
$ and $\mathbf{x}^{\prime }$ have no relative motion. \ Then $\partial
_{x_{i}}u_{j}^{\prime }=0$, $\partial _{x_{i}^{\prime }}u_{j}=0$, etc., and $%
\partial _{t}$ is performed with both $\mathbf{x}$ and $\mathbf{x}^{\prime }$
fixed. \ The change of independent variables from $\mathbf{x}$ and $\mathbf{x%
}^{\prime }$ to the sum and difference independent variables is: 
\begin{equation}
\mathbf{X}\equiv \left( \mathbf{x}+\mathbf{x}^{\prime }\right) /2\text{ \
and \ }\mathbf{r}\equiv \mathbf{x}-\mathbf{x}^{\prime },\ \ \text{and define 
}r\equiv \left\vert \mathbf{r}\right\vert .  \label{change}
\end{equation}%
The relationship between the partial derivatives is 
\begin{equation}
\partial _{x_{i}}=\partial _{r_{i}}+\frac{1}{2}\partial _{X_{i}}\text{ , }\
\partial _{x_{i}^{\prime }}=-\partial _{r_{i}}+\frac{1}{2}\partial _{X_{i}}%
\text{ \ , }\partial _{X_{i}}=\partial _{x_{i}}+\partial _{x_{i}^{\prime }}%
\text{ \ , }\partial _{r_{i}}=\frac{1}{2}\left( \partial _{x_{i}}-\partial
_{x_{i}^{\prime }}\right) .  \label{1identderivs}
\end{equation}%
The change of variables organizes the equations in a revealing way because
of the following properties. \ In the case of homogeneous turbulence, $%
\partial _{X_{i}}$ operating on a statistic produces zero because that
derivative is the rate of change with respect to the place where the
measurement is performed. \ Consider a term in an equation composed of $%
\partial _{X_{i}}$ operating on a statistic. \ For locally homogeneous
turbulence, that term becomes negligible as $r$ is decreased relative to the
integral scale. \ For the homogeneous and locally homogeneous cases, the
statistical equations retain their dependence on $\mathbf{r}$, which is the
displacement vector of two points of measurement. \ Subtracting (\ref{1NSE})
at $\mathbf{x}^{\prime }$ from (\ref{1NSE}) at $\mathbf{x}$, performing the
change of variables (\ref{change}), and using (\ref{1identderivs}) gives 
\begin{eqnarray}
\partial _{t}v_{i}+U_{n}\partial _{X_{n}}v_{i}+v_{n}\partial _{r_{n}}v_{i}
&=&-P_{i}+\nu \left( \partial _{x_{n}}\partial _{x_{n}}v_{i}+\partial
_{x_{n}^{\prime }}\partial _{x_{n}^{\prime }}v_{i}\right) ,  \label{2ndstep}
\\
\text{where }v_{i} &\equiv &u_{i}-u_{i}^{\prime },U_{n}\equiv \left(
u_{n}+u_{n}^{\prime }\right) /2,  \notag \\
\text{and }P_{i} &\equiv &\partial _{x_{i}}p-\partial _{x_{i}^{\prime
}}p^{\prime }.
\end{eqnarray}

\qquad Now multiply (\ref{2ndstep}) by the product $v_{j}v_{k}\cdot \cdot
\cdot v_{l}$ which contains $N-1$ factors of velocity difference, each
factor having a distinct index. \ Sum the $N$ equations as required to
produce symmetry under interchange of each pair of indices, excluding the
summation index $n$. \ French braces, i.e., $\left\{ \circ \right\} $,
denote the sum of all terms of a given type that produce symmetry under
interchange of each pair of indices. \ The differentiation chain rule gives 
\begin{eqnarray}
\left\{ v_{j}v_{k}\cdot \cdot \cdot v_{l}\partial _{t}v_{i}\right\}
&=&\partial _{t}\left( v_{j}v_{k}\cdot \cdot \cdot v_{l}v_{i}\right) ,
\label{tempderiv} \\
\left\{ v_{j}v_{k}\cdot \cdot \cdot v_{l}U_{n}\partial _{X_{n}}v_{i}\right\}
&=&U_{n}\partial _{X_{n}}\left( v_{j}v_{k}\cdot \cdot \cdot
v_{l}v_{i}\right) =\partial _{X_{n}}\left( U_{n}v_{j}v_{k}\cdot \cdot \cdot
v_{l}v_{i}\right) ,  \label{1p} \\
\left\{ v_{j}v_{k}\cdot \cdot \cdot v_{l}v_{n}\partial _{r_{n}}v_{i}\right\}
&=&v_{n}\partial _{r_{n}}\left( v_{j}v_{k}\cdot \cdot \cdot
v_{l}v_{i}\right) =\partial _{r_{n}}\left( v_{n}v_{j}v_{k}\cdot \cdot \cdot
v_{l}v_{i}\right) .  \label{1pa}
\end{eqnarray}
The right-most expressions in (\ref{1p}) and (\ref{1pa}) follow from the
incompressibility property obtained from (\ref{1identderivs}) and the fact
that $\partial _{x_{i}}u_{j}^{\prime }=0$, $\partial _{x_{i}^{\prime
}}u_{j}=0$; namely, $\partial _{X_{n}}U_{n}=0,\partial
_{X_{n}}v_{n}=0,\partial _{r_{n}}U_{n}=0,\partial _{r_{n}}v_{n}=0$. \ The
viscous term in (\ref{2ndstep}) produces $\nu \left\{ v_{j}v_{k}\cdot \cdot
\cdot v_{l}\left( \partial _{x_{n}}\partial _{x_{n}}v_{i}+\partial
_{x_{n}^{\prime }}\partial _{x_{n}^{\prime }}v_{i}\right) \right\} $; this
expression is treated in the Archive. \ Thereby 
\begin{eqnarray}
&&\partial _{t}\left( v_{j}\cdot \cdot \cdot v_{i}\right) +\partial
_{X_{n}}\left( U_{n}v_{j}\cdot \cdot \cdot v_{i}\right) +\partial
_{r_{n}}\left( v_{n}v_{j}\cdot \cdot \cdot v_{i}\right) =  \notag \\
&&-\left\{ v_{j}\cdot \cdot \cdot v_{l}P_{i}\right\} +2\nu \left[ \left(
\partial _{r_{n}}\partial _{r_{n}}+\frac{1}{4}\partial _{X_{n}}\partial
_{X_{n}}\right) \left( v_{j}\cdot \cdot \cdot v_{i}\right) -\left\{
v_{k}\cdot \cdot \cdot v_{l}e_{ij}\right\} \right] ,  \label{0exactunave}
\end{eqnarray}
\begin{equation}
\text{where }e_{ij}\equiv \left( \partial _{x_{n}}u_{i}\right) \left(
\partial _{x_{n}}u_{j}\right) +\left( \partial _{x_{n}^{\prime
}}u_{i}^{\prime }\right) \left( \partial _{x_{n}^{\prime }}u_{j}^{\prime
}\right) =\left( \partial _{x_{n}}v_{i}\right) \left( \partial
_{x_{n}}v_{j}\right) +\left( \partial _{x_{n}^{\prime }}v_{i}^{\prime
}\right) \left( \partial _{x_{n}^{\prime }}v_{j}^{\prime }\right) .
\label{eij}
\end{equation}

\qquad The quantity $\left\{ v_{j}\cdot \cdot \cdot v_{l}P_{i}\right\} $\
can be expressed differently on the basis that (\ref{1identderivs}) allows $%
P_{i}$\ to be written as $P_{i}=\partial _{X_{i}}\left( p-p^{\prime }\right) 
$. The derivation is in the Archive; the alternative formula is 
\begin{equation}
\left\{ v_{j}v_{k}\cdot \cdot \cdot v_{l}P_{i}\right\} =\left\{ \partial
_{X_{i}}\left[ v_{j}v_{k}\cdot \cdot \cdot v_{l}\left( p-p^{\prime }\right) %
\right] \right\} -2\left( N-1\right) \left( p-p^{\prime }\right) \left\{
\left( s_{ij}-s_{ij}^{\prime }\right) v_{k}\cdot \cdot \cdot v_{l}\right\} ,
\label{1Palternative}
\end{equation}%
where the rate of strain tensor $s_{ij}$\ is defined by $s_{ij}\equiv \left(
\partial _{x_{i}}u_{j}+\partial _{x_{j}}u_{i}\right) /2.$

\subsection{\textbf{Hierarchy of exact statistical equations}}

\qquad Consider the ensemble average because it commutes with temporal and
spatial derivatives. \ The above notation of explicit indices is burdensome.
\ Because the tensors are symmetric, it suffices to show only the number of
indices. \ Define the following statistical tensors which are symmetric
under interchange of any pair of indices, excluding the summation index $n$
in the definition of $\mathbf{F}_{\left[ N+1\right] }$: 
\begin{equation}
\mathbf{D}_{\left[ N\right] }\equiv \left\langle v_{j}\cdot \cdot \cdot
v_{i}\right\rangle ,\mathbf{F}_{\left[ N+1\right] }\equiv \left\langle
U_{n}v_{j}\cdot \cdot \cdot v_{i}\right\rangle ,\mathbf{T}_{\left[ N\right]
}\equiv \left\langle \left\{ v_{j}\cdot \cdot \cdot v_{l}P_{i}\right\}
\right\rangle ,\mathbf{E}_{\left[ N\right] }\equiv \left\langle \left\{
v_{k}\cdot \cdot \cdot v_{l}e_{ij}\right\} \right\rangle ,
\label{1implicitindex}
\end{equation}%
where angle brackets $\left\langle {}\right\rangle $ denote the ensemble
average, and the subscripts $N$\ and $N+1$ within square brackets denote the
number of indices. \ The left-hand side of each definition in (\ref%
{1implicitindex}) is in implicit-index notation for which only the number of
indices is given; the right-hand sides in (\ref{1implicitindex}) are in
explicit-index notation. \ The argument list for each tensor is understood
to be $\left( \mathbf{X},\mathbf{r},t\right) $. \ The ensemble average of (%
\ref{0exactunave}) is 
\begin{equation}
\partial _{t}\mathbf{D}_{\left[ N\right] }+\nabla _{\mathbf{X}}\bullet 
\mathbf{F}_{\left[ N+1\right] }+\nabla _{\mathbf{r}}\bullet \mathbf{D}_{%
\left[ N+1\right] }=-\mathbf{T}_{\left[ N\right] }+2\nu \left[ \left( \nabla
_{\mathbf{r}}^{2}+\frac{1}{4}\nabla _{\mathbf{X}}^{2}\right) \mathbf{D}_{%
\left[ N\right] }-\mathbf{E}_{\left[ N\right] }\right] ,
\label{2implicitindex}
\end{equation}%
where, $\nabla _{\mathbf{X}}\bullet \mathbf{F}_{\left[ N+1\right] }\equiv
\partial _{X_{n}}\left\langle U_{n}v_{j}\cdot \cdot \cdot v_{i}\right\rangle
,\nabla _{\mathbf{r}}\bullet \mathbf{D}_{\left[ N+1\right] }\equiv \partial
_{r_{n}}\left\langle v_{n}v_{j}\cdot \cdot \cdot v_{i}\right\rangle ,\nabla
_{\mathbf{r}}^{2}\equiv \partial _{r_{n}}\partial _{r_{n}},\nabla _{\mathbf{X%
}}^{2}\equiv \partial _{X_{n}}\partial _{X_{n}}$. \ The notations $\nabla _{%
\mathbf{X}}\bullet $, $\nabla _{\mathbf{X}}^{2}$, $\nabla _{\mathbf{r}%
}\bullet $, and $\nabla _{\mathbf{r}}^{2}$ are the divergence and Laplacian
operators in $\mathbf{X}$-space and $\mathbf{r}$-space, respectively.

\section{\textbf{Homogeneous and locally homogeneous turbulence}}

\qquad Consider homogeneous turbulence and locally homogeneous turbulence;
the latter applies for small $r$ and large Reynolds number. \ The variation
of the statistics with the location of measurement or of evaluation is zero
for the homogeneous case and is neglected for the locally homogeneous case.
\ Since that location is $\mathbf{X}$\textbf{,} the result of $\nabla _{%
\mathbf{X}}$ operating on a statistic vanishes or is neglected as the case
may be. \ Thus the terms $\nabla _{\mathbf{X}}\bullet \mathbf{F}_{\left[ N+1%
\right] }$\ and $\frac{1}{4}\nabla _{\mathbf{X}}^{2}\mathbf{D}_{\left[ N%
\right] }$ are deleted in (\ref{2implicitindex}); then (\ref{2implicitindex}%
) becomes, 
\begin{equation}
\partial _{t}\mathbf{D}_{\left[ N\right] }+\nabla _{\mathbf{r}}\bullet 
\mathbf{D}_{\left[ N+1\right] }=-\mathbf{T}_{\left[ N\right] }+2\nu \left[
\nabla _{\mathbf{r}}^{2}\mathbf{D}_{\left[ N\right] }-\mathbf{E}_{\left[ N%
\right] }\right] .  \label{1iso}
\end{equation}%
Because the $\mathbf{X}$-dependence is deleted, the argument list for each
tensor is understood to be $\left( \mathbf{r},t\right) $. \ Note that $%
\partial _{t}\mathbf{D}_{\left[ N\right] }$\ is not necessarily negligible
for homogeneous turbulence. \ The ensemble average of (\ref{1Palternative})
contains $\left\langle \left\{ \partial _{X_{i}}\left[ v_{j}v_{k}\cdot \cdot
\cdot v_{l}\left( p-p^{\prime }\right) \right] \right\} \right\rangle
=\left\{ \partial _{X_{i}}\left\langle v_{j}v_{k}\cdot \cdot \cdot
v_{l}\left( p-p^{\prime }\right) \right\rangle \right\} =\left\{ 0\right\}
=0 $. \ Thus, (\ref{1Palternative}) gives the alternative that 
\begin{equation}
\mathbf{T}_{\left[ N\right] }=-2\left( N-1\right) \left\langle \left(
p-p^{\prime }\right) \left\{ \left( s_{ij}-s_{ij}^{\prime }\right)
v_{k}\cdot \cdot \cdot v_{l}\right\} \right\rangle .
\end{equation}%
One distinction between (\ref{1iso}) and the hierarchy equations given in
equations (13) and (17) by Arad \textit{et al.} (1999) is that their $t$-
and $\mathbf{r}$-derivatives operate on only one velocity difference within
their product of such differences, whereas the derivatives in (\ref%
{0exactunave}) and thus in (\ref{1iso}) operate on all $N$ of the velocity
differences.

\section{\textbf{Isotropic and locally isotropic turbulence}}

\qquad Consider isotropic turbulence and locally isotropic turbulence; the
latter applies for small $r$ and large Reynolds number. \ Locally isotropic
flows are a subset of locally homogeneous flows (Monin \& Yaglom Sec. 13.3,
1975) and similarly for the relationship between isotropic and homogeneous
flows. \ Thus, the dynamical equations for locally isotropic and isotropic
turbulence are obtained from (\ref{1iso}) such that the variable $\mathbf{X}$%
\ and the terms $\nabla _{\mathbf{X}}\bullet \mathbf{F}_{\left[ N+1\right] }$%
\ and $\frac{1}{4}\nabla _{\mathbf{X}}^{2}\mathbf{D}_{\left[ N\right] }$
(see \ref{2implicitindex}) do not appear. \ The tensors $\mathbf{D}_{\left[ N%
\right] }$, $\mathbf{T}_{\left[ N\right] }$, and $\mathbf{E}_{\left[ N\right]
}$\ in (\ref{1implicitindex}) obey the isotropic formula. \ The Kronecker
delta $\delta _{ij}$ is defined by $\delta _{ij}=1$ if $i=j$ and $\delta
_{ij}=0$ if $i\neq j$. $\ $\ Let $\mathbf{\delta }_{\left[ 2P\right] }$
denote the product of $P$ Kronecker deltas having $2P$ distinct indices, and
let $\mathbf{W}_{\left[ N\right] }\left( \mathbf{r}\right) $ denote the
product of $N$ factors $\frac{r_{i}}{r}$ each with a distinct index; the
argument $\mathbf{r}$\ is omitted when clarity does not suffer. \ Because
each tensor in (\ref{1implicitindex}) is symmetric under interchange of any
two indices, their isotropic formulas are particularly simple. \ Each
formula is a sum of $M+1$\ terms where $M=N/2$ if $N$ is even, and $M=\left(
N-1\right) /2$ if $N$ is odd. \ Each term is the product of a distinct
scalar function with a $\mathbf{W}_{\left[ N\right] }$\ and a $\mathbf{%
\delta }_{\left[ 2P\right] }$. \ From one term to the next a pair of indices
is transferred from a $\mathbf{W}_{\left[ N\right] }$ to a $\mathbf{\delta }%
_{\left[ 2P\right] }$; examples are in the Archive. \ For the tensor $%
\mathbf{D}_{\left[ N\right] }$, denote the $P$th scalar function by $%
D_{N,P}\left( r,t\right) $. \ The isotropic formula for $\mathbf{D}_{\left[ N%
\right] }$\ is 
\begin{equation}
\mathbf{D}_{\left[ N\right] }\left( \mathbf{r},t\right) =\overset{M}{%
\underset{P=0}{\sum }}D_{N,P}\left( r,t\right) \left\{ \mathbf{W}_{\left[
N-2P\right] }\left( \mathbf{r}\right) \mathbf{\delta }_{\left[ 2P\right]
}\right\} ,  \label{1isoformula}
\end{equation}
and the isotropic formulas for $\mathbf{T}_{\left[ N\right] }$ and $\mathbf{E%
}_{\left[ N\right] }$ have the analogous notation. \ Recall from Sec. 2 the
meaning of the notation $\left\{ \circ \right\} $ whereby $\left\{ \mathbf{W}%
_{\left[ N-2P\right] }\left( \mathbf{r}\right) \mathbf{\delta }_{\left[ 2P%
\right] }\right\} $ denotes the sum of all terms of the type $\mathbf{W}_{%
\left[ N-2P\right] }\left( \mathbf{r}\right) \mathbf{\delta }_{\left[ 2P%
\right] }$ that produce symmetry under interchange of each pair of indices.
An example is $\left\{ \mathbf{W}_{\left[ 1\right] }\left( \mathbf{r}\right) 
\mathbf{\delta }_{\left[ 2\right] }\right\} =\frac{r_{k}}{r}\delta _{ij}+%
\frac{r_{j}}{r}\delta _{ki}+\frac{r_{i}}{r}\delta _{jk}$.

\qquad A special Cartesian coordinate system simplifies the isotropic
formulas. \ This coordinate system has the positive $1$-axis parallel to the
direction of $\mathbf{r}$, and the $2$- and $3$-axes are therefore
perpendicular to $\mathbf{r}$. \ Let $N_{1}$, $N_{2}$, and $N_{3}$ be the
number of indices of a component of $\mathbf{D}_{\left[ N\right] }$ that are 
$1$, $2$, and $3$, respectively; such that $N=N_{1}+N_{2}+N_{3}$. \ Because
of symmetry, the order of indices is immaterial so that a component of $%
\mathbf{D}_{\left[ N\right] }$ can be identified by $N_{1}$, $N_{2}$, and $%
N_{3}$. \ Thus, denote a component of $\mathbf{D}_{\left[ N\right] }$ by $D_{%
\left[ N_{1},N_{2},N_{3}\right] }$ which is a function of $\mathbf{r}$ and $%
t $. \ Likewise, $\left\{ \mathbf{W}_{\left[ N-2P\right] }\left( \mathbf{r}%
\right) \mathbf{\delta }_{\left[ 2P\right] }\right\} _{\left[
N_{1},N_{2},N_{3}\right] }$\ is a specific component of the tensor $\left\{ 
\mathbf{W}_{\left[ N-2P\right] }\left( \mathbf{r}\right) \mathbf{\delta }_{%
\left[ 2P\right] }\right\} $. \ If, in (\ref{1isoformula}) $N_{1}$ of the
indices are assigned the value $1$, and $N_{2}$ and $N_{3}$ of the indices
are assigned the values $2$ and $3$, respectively, then $D_{\left[
N_{1},N_{2},N_{3}\right] }$\ and $\left\{ \mathbf{W}_{\left[ N-2P\right]
}\left( \mathbf{r}\right) \mathbf{\delta }_{\left[ 2P\right] }\right\} _{%
\left[ N_{1},N_{2},N_{3}\right] }$\ will appear on the left-hand and
right-hand sides of (\ref{1isoformula}), respectively. \ The $\left\{ 
\mathbf{W}_{\left[ N-2P\right] }\left( \mathbf{r}\right) \mathbf{\delta }_{%
\left[ 2P\right] }\right\} _{\left[ N_{1},N_{2},N_{3}\right] }$ are
numerical coefficients that do not depend on $\mathbf{r}$ because $\frac{%
r_{1}}{r}=\frac{r}{r}=1$, $\frac{r_{2}}{r}=\frac{r_{3}}{r}=0$. \ From the
Archive, the values of the coefficients are: 
\begin{equation*}
\text{ \ }
\end{equation*}

\begin{equation}
\text{if }2P<N_{2}+N_{3}\text{ then }\left\{ \mathbf{W}_{\left[ N-2P\right]
}\left( \mathbf{r}\right) \mathbf{\delta }_{\left[ 2P\right] }\right\} _{%
\left[ N_{1},N_{2},N_{3}\right] }=0,\text{ otherwise,}  \label{0isocoefs}
\end{equation}
\begin{equation}
\left\{ \mathbf{W}_{\left[ N-2P\right] }\left( \mathbf{r}\right) \mathbf{%
\delta }_{\left[ 2P\right] }\right\} _{\left[ N_{1},N_{2},N_{3}\right]
}=N_{1}!N_{2}!N_{3}!/\left[ \left( N-2P\right) !2^{P}\left( \frac{N_{2}}{2}%
\right) !\left( \frac{N_{3}}{2}\right) !\left( P-\frac{N_{2}}{2}-\frac{N_{3}%
}{2}\right) !\right] .  \label{1isocoefs}
\end{equation}

\qquad By applying (\ref{1isoformula}-\ref{1isocoefs}) for all combinations
of indices, one can determine which components $D_{\left[ N_{1},N_{2},N_{3}%
\right] }$\ are zero and which are nonzero, identify $M+1$\ linearly
independent equations that determine\ the $D_{N,P}$ in terms of $M+1$ of the 
$D_{\left[ N_{1},N_{2},N_{3}\right] }$,\ and find algebraic relationships
between the remaining nonzero $D_{\left[ N_{1},N_{2},N_{3}\right] }$. \ The
derivations are in the Archive; a summary follows.

\qquad A component $D_{\left[ N_{1},N_{2},N_{3}\right] }$ is nonzero only if
both $N_{2}$ and $N_{3}$ are even, and therefore $N_{1}$ is odd if $N$ is
odd, and $N_{1}$ is even if $N$ is even. \ Thereby, $\left( M+1\right)
\left( M+2\right) /2$\ components are nonzero. \ There are $3^{N}$
components of $\mathbf{D}_{\left[ N\right] }$; thus the other $3^{N}-\left(
M+1\right) \left( M+2\right) /2$ components are zero.

\qquad There exist $\left( M+1\right) M/2$\ kinematic relationships among
the nonzero components of $\mathbf{D}_{\left[ N\right] }$. \ For each of the 
$M+1$ cases of $N_{1}$, these relationships are expressed by the
proportionality 
\begin{eqnarray*}
D_{\left[ N_{1},2L,0\right] } &:&D_{\left[ N_{1},2L-2,2\right] }:D_{\left[
N_{1},2L-4,4\right] }:\cdot \cdot \cdot :D_{\left[ N_{1},0,2L\right] }= \\
\left[ \left( 2L\right) !0!/L!0!\right] &:&\left[ \left( 2L-2\right)
!2!/\left( L-1\right) !1!\right] :\left[ \left( 2L-4\right) !4!/\left(
L-2\right) !2!\right] :\cdot \cdot \cdot :\left[ 0!\left( 2L\right) !/0!L!%
\right] .
\end{eqnarray*}
\begin{equation}
\text{ \ }  \label{kinematic}
\end{equation}
For $N=4$ with $L=2$,\ (\ref{kinematic}) gives $D_{\left[ 0,4,0\right] }:D_{%
\left[ 0,2,2\right] }:D_{\left[ 0,0,4\right] }=12:4:12$. \ In explicit-index
notation this can be written as $D_{2222}=3D_{2233}=D_{3333}$, which was
discovered by Millionshtchikov (1941) and is the only previously known such
relationship. \ Now, all such relationships are known from (\ref{kinematic}).

\qquad There remain $M+1$\ linearly independent nonzero components of $%
\mathbf{D}_{\left[ N\right] }$. \ This must be so because there are $M+1$\
terms\ in (\ref{1isoformula}) and the $M+1$ scalar functions $D_{N,P}$\
therein must be related to $M+1$\ components. \ Consider the $M+1$\ linearly
independent equations that determine\ the $D_{N,P}$ in terms of $M+1$ of the 
$D_{\left[ N_{1},N_{2},N_{3}\right] }$. \ For simplicity, the chosen
components can all have $N_{3}=0$; i.e., the choice of linearly independent
components can be $D_{\left[ N,0,0\right] }$, $D_{\left[ N-2,2,0\right] }$, $%
D_{\left[ N-4,4,0\right] }$, $\cdot \cdot \cdot $, $D_{\left[ N-2M,2M,0%
\right] }$. \ As described above, assigning index values in (\ref%
{1isoformula}) results in the chosen components on the left-hand side and
algebraic expressions on the right-hand side that contain the coefficients (%
\ref{0isocoefs}-\ref{1isocoefs}). \ In the Archive, those equations are
expressed in matrix form and solved by matrix inversion methods. \ Given
experimental or DNS data or a theoretical formula for the chosen components,
the solution of the algebraic equations determines the functions $D_{N,P}$\
in (\ref{1isoformula}); then (\ref{1isoformula}) completely specifies the
tensor $\mathbf{D}_{\left[ N\right] }$.

\qquad The matrix algorithm in the Archive is an efficient means of
determining isotropic expressions for the terms $\nabla _{\mathbf{r}}\bullet 
\mathbf{D}_{\left[ N+1\right] }$\ and $\nabla _{\mathbf{r}}^{2}\mathbf{D}_{%
\left[ N\right] }$\ in (\ref{1iso}). \ From the example for $N=2$ in the
Archive, (\ref{1iso}) gives the two scalar equations \pagebreak 
\begin{eqnarray}
\partial _{t}D_{11}+\left( \partial _{r}+\frac{2}{r}\right) D_{111}-\frac{4}{%
r}D_{122} &=&-T_{11}+2\nu \left[ \left( \partial _{r}^{2}+\frac{2}{r}%
\partial _{r}-\frac{4}{r^{2}}\right) D_{11}+\frac{4}{r^{2}}D_{22}\right]
-2\nu E_{11}  \notag \\
&=&2\nu \left[ \partial _{r}^{2}D_{11}+\frac{2}{r}\partial _{r}D_{11}+\frac{4%
}{r^{2}}\left( D_{22}-D_{11}\right) \right] -4\varepsilon /3,
\label{2ndorder}
\end{eqnarray}%
\begin{eqnarray}
\partial _{t}D_{22}+\left( \partial _{r}+\frac{4}{r}\right) D_{122}
&=&-T_{22}+2\nu \left[ \frac{2}{r^{2}}D_{11}+\left( \partial _{r}^{2}+\frac{2%
}{r}\partial _{r}-\frac{2}{r^{2}}\right) D_{22}\right] -2\nu E_{22}  \notag
\\
&=&2\nu \left[ \partial _{r}^{2}D_{22}+\frac{2}{r}\partial _{r}D_{22}-\frac{2%
}{r^{2}}\left( D_{22}-D_{11}\right) \right] -4\varepsilon /3,
\label{2ndorder2}
\end{eqnarray}%
where use was made of the fact (Hill 1997a) that local isotropy gives $%
T_{11}=T_{22}=0$ and $2\nu E_{11}=2\nu E_{22}=4\varepsilon /3$, where $%
\varepsilon $\ is the average energy dissipation rate. \ Since (\ref%
{2ndorder}-\ref{2ndorder2}) are the same as equations (43-44) of Hill
(1997a), and since Hill (1997a) shows how these equations lead to
Kolmogorov's equation and his 4/5 law, further discussion of (\ref{2ndorder}-%
\ref{2ndorder2})\ is unnecessary. \ From the example for $N=3$ in the
Archive, 
\begin{eqnarray}
\partial _{t}D_{111}+\left( \partial _{r}+\frac{2}{r}\right) D_{1111}-\frac{6%
}{r}D_{1122} &=&-T_{111}+2\nu \left[ \left( \nabla _{\mathbf{r}}^{2}\mathbf{D%
}\right) _{111}-E_{111}\right] ,  \label{HillBor1} \\
\partial _{t}D_{122}+\left( \partial _{r}+\frac{4}{r}\right) D_{1122}-\frac{4%
}{3r}D_{2222} &=&-T_{122}+2\nu \left[ \left( \nabla _{\mathbf{r}}^{2}\mathbf{%
D}\right) _{122}-E_{122}\right] ,  \label{HillBor2}
\end{eqnarray}%
\begin{eqnarray}
\left( \nabla _{\mathbf{r}}^{2}\mathbf{D}\right) _{111} &\equiv &\left(
\partial _{r}^{2}+\frac{2}{r}\partial _{r}-\frac{6}{r^{2}}\right) D_{111}+%
\frac{12}{r^{2}}D_{122}  \label{HillBor3} \\
&=&\left( -\frac{4}{r^{2}}+\frac{4}{r}\partial _{r}+\partial _{r}^{2}\right)
D_{111},  \notag \\
\left( \nabla _{\mathbf{r}}^{2}\mathbf{D}\right) _{122} &\equiv &\frac{2}{%
r^{2}}D_{111}+\left( \partial _{r}^{2}+\frac{2}{r}\partial _{r}-\frac{8}{%
r^{2}}\right) D_{122}  \notag \\
&=&\frac{1}{6}\left( \frac{4}{r^{2}}-\frac{4}{r}\partial _{r}+5\partial
_{r}^{2}+r\partial _{r}^{3}\right) D_{111}.  \label{HillBor4}
\end{eqnarray}%
The incompressibility condition, $D_{122}=\frac{1}{6}\left(
D_{111}+r\partial _{r}D_{111}\right) $, was used in (\ref{HillBor3}-\ref%
{HillBor4}). \ Since Hill \& Boratav (2001)\ discuss these equations and
evaluate them using data, further discussion of \ (\ref{HillBor1}-\ref%
{HillBor4})\ is unnecessary.

\qquad The terms $\partial _{t}\mathbf{D}_{\left[ N\right] }$, $-\mathbf{T}_{%
\left[ N\right] }$, and $-2\nu \mathbf{E}_{\left[ N\right] }$\ in (\ref{1iso}%
) have a repetitive structure in the isotropic equations; e.g., for $N=4$
the 3 equations are 
\begin{eqnarray}
\partial _{t}D_{1111}+\left( \nabla _{\mathbf{r}}\bullet \mathbf{D}_{\left[ 5%
\right] }\right) _{1111} &=&-T_{1111}+2\nu \left[ \left( \nabla _{\mathbf{r}%
}^{2}\mathbf{D}_{\left[ 4\right] }\right) _{1111}-E_{1111}\right] , \\
\partial _{t}D_{1122}+\left( \nabla _{\mathbf{r}}\bullet \mathbf{D}_{\left[ 5%
\right] }\right) _{1122} &=&-T_{1122}+2\nu \left[ \left( \nabla _{\mathbf{r}%
}^{2}\mathbf{D}_{[4]}\right) _{1122}-E_{1122}\right] , \\
\partial _{t}D_{2222}+\left( \nabla _{\mathbf{r}}\bullet \mathbf{D}_{\left[ 5%
\right] }\right) _{2222} &=&-T_{2222}+2\nu \left[ \left( \nabla _{\mathbf{r}%
}^{2}\mathbf{D}_{[4]}\right) _{2222}-E_{2222}\right] .
\end{eqnarray}%
Thus, it suffices to give the isotropic formulas for the divergence $\nabla
_{\mathbf{r}}\bullet \mathbf{D}_{\left[ N+1\right] }$\ and Laplacian $\nabla
_{\mathbf{r}}^{2}\mathbf{D}_{\left[ N\right] }$. Table 1 gives those
isotropic formulas for $N=4$ to $8$. \ The case $N=8$\ was not given by Hill
(2001). \ For $N=4$ and $5$\ there are $M+1=3$ equations; there are $M+1=4$
equations for both $N=6$ and $7$; there are $M+1=5$\ equations for $N=8$.

------------------------------------------------------------------------------------------------------%
\begin{equation*}
N=4
\end{equation*}%
\bigskip $%
\begin{array}{c}
\left( \partial _{r}+\frac{2}{r}\right) D_{\left[ 5,0,0\right] }-\frac{8}{r}%
D_{\left[ 3,2,0\right] } \\ 
\left( \partial _{r}+\frac{4}{r}\right) D_{\left[ 3,2,0\right] }-\frac{8}{3r}%
D_{\left[ 1,4,0\right] } \\ 
\left( \partial _{r}+\frac{6}{r}\right) D_{\left[ 1,4,0\right] }%
\end{array}%
$ $\ \ \ \ \ 
\begin{array}{c}
\left( \partial _{r}^{2}+\frac{2}{r}\partial _{r}-\frac{8}{r^{2}}\right) D_{%
\left[ 4,0,0\right] }+\frac{24}{r^{2}}D_{\left[ 2,2,0\right] } \\ 
\left( \partial _{r}^{2}+\frac{2}{r}\partial _{r}-\frac{14}{r^{2}}\right) D_{%
\left[ 2,2,0\right] }+\frac{2}{r^{2}}D_{\left[ 4,0,0\right] }+\frac{8}{3r^{2}%
}D_{\left[ 0,4,0\right] } \\ 
\left( \partial _{r}^{2}+\frac{2}{r}\partial _{r}-\frac{4}{r^{2}}\right) D_{%
\left[ 0,4,0\right] }+\frac{12}{r^{2}}D_{\left[ 2,2,0\right] }%
\end{array}%
$

------------------------------------------------------------------------------------------------------%
\begin{equation*}
N=5
\end{equation*}%
$%
\begin{array}{c}
\left( \partial _{r}+\frac{2}{r}\right) D_{\left[ 6,0,0\right] }-\frac{10}{r}%
D_{\left[ 4,2,0\right] } \\ 
\left( \partial _{r}+\frac{4}{r}\right) D_{\left[ 4,2,0\right] }-\frac{4}{r}%
D_{\left[ 2,4,0\right] } \\ 
\left( \partial _{r}+\frac{6}{r}\right) D_{\left[ 2,4,0\right] }-\frac{6}{5r}%
D_{\left[ 0,6,0\right] }%
\end{array}%
$ \ \ \ \ \ \ $%
\begin{array}{c}
\left( \partial _{r}^{2}+\frac{2}{r}\partial _{r}-\frac{10}{r^{2}}\right) D_{%
\left[ 5,0,0\right] }+\frac{40}{r^{2}}D_{\left[ 3,2,0\right] } \\ 
\left( \partial _{r}^{2}+\frac{2}{r}\partial _{r}-\frac{20}{r^{2}}\right) D_{%
\left[ 3,2,0\right] }+\frac{2}{r^{2}}D_{\left[ 5,0,0\right] }+\frac{8}{r^{2}}%
D_{\left[ 1,4,0\right] } \\ 
\left( \partial _{r}^{2}+\frac{2}{r}\partial _{r}-\frac{14}{r^{2}}\right) D_{%
\left[ 1,4,0\right] }+\frac{12}{r^{2}}D_{\left[ 3,2,0\right] }%
\end{array}%
$

------------------------------------------------------------------------------------------------------%
\begin{equation*}
N=6
\end{equation*}%
$%
\begin{array}{c}
\left( \partial _{r}+\frac{2}{r}\right) D_{\left[ 7,0,0\right] }-\frac{12}{r}%
D_{\left[ 5,2,0\right] } \\ 
\left( \partial _{r}+\frac{4}{r}\right) D_{\left[ 5,2,0\right] }-\frac{16}{3r%
}D_{\left[ 3,4,0\right] } \\ 
\left( \partial _{r}+\frac{6}{r}\right) D_{\left[ 3,4,0\right] }-\frac{12}{5r%
}D_{\left[ 1,6,0\right] } \\ 
\left( \partial _{r}+\frac{8}{r}\right) D_{\left[ 1,6,0\right] }%
\end{array}%
$ \ \ \ \ \ $%
\begin{array}{c}
\left( \partial _{r}^{2}+\frac{2}{r}\partial _{r}-\frac{12}{r^{2}}\right) D_{%
\left[ 6,0,0\right] }+\frac{60}{r^{2}}D_{\left[ 4,2,0\right] } \\ 
\left( \partial _{r}^{2}+\frac{2}{r}\partial _{r}-\frac{26}{r^{2}}\right) D_{%
\left[ 4,2,0\right] }+\frac{2}{r^{2}}D_{\left[ 6,0,0\right] }+\frac{16}{r^{2}%
}D_{\left[ 2,4,0\right] } \\ 
\left( \partial _{r}^{2}+\frac{2}{r}\partial _{r}-\frac{24}{r^{2}}\right) D_{%
\left[ 2,4,0\right] }+\frac{12}{r^{2}}D_{\left[ 4,2,0\right] }+\frac{12}{%
5r^{2}}D_{\left[ 0,6,0\right] } \\ 
\left( \partial _{r}^{2}+\frac{2}{r}\partial _{r}-\frac{6}{r^{2}}\right) D_{%
\left[ 0,6,0\right] }+\frac{30}{r^{2}}D_{\left[ 2,4,0\right] }%
\end{array}%
$

------------------------------------------------------------------------------------------------------%
\begin{equation*}
N=7
\end{equation*}

\bigskip $%
\begin{array}{c}
\left( \partial _{r}+\frac{2}{r}\right) D_{\left[ 8,0,0\right] }-\frac{14}{r}%
D_{\left[ 6,2,0\right] } \\ 
\left( \partial _{r}+\frac{4}{r}\right) D_{\left[ 6,2,0\right] }-\frac{20}{3r%
}D_{\left[ 4,4,0\right] } \\ 
\left( \partial _{r}+\frac{6}{r}\right) D_{\left[ 4,4,0\right] }-\frac{18}{5r%
}D_{\left[ 2,6,0\right] } \\ 
\left( \partial _{r}+\frac{8}{r}\right) D_{\left[ 2,6,0\right] }-\frac{8}{7r}%
D_{\left[ 0,8,0\right] }%
\end{array}%
$\ \ $%
\begin{array}{c}
\left( \partial _{r}^{2}+\frac{2}{r}\partial _{r}-\frac{14}{r^{2}}\right) D_{%
\left[ 7,0,0\right] }+\frac{84}{r^{2}}D_{\left[ 5,2,0\right] } \\ 
\left( \partial _{r}^{2}+\frac{2}{r}\partial _{r}-\frac{32}{r^{2}}\right) D_{%
\left[ 5,2,0\right] }+\frac{2}{r^{2}}D_{\left[ 7,0,0\right] }+\frac{80}{%
3r^{2}}D_{\left[ 3,4,0\right] } \\ 
\left( \partial _{r}^{2}+\frac{2}{r}\partial _{r}-\frac{34}{r^{2}}\right) D_{%
\left[ 3,4,0\right] }+\frac{12}{r^{2}}D_{\left[ 5,2,0\right] }+\frac{36}{%
5r^{2}}D_{\left[ 1,6,0\right] } \\ 
\left( \partial _{r}^{2}+\frac{2}{r}\partial _{r}-\frac{20}{r^{2}}\right) D_{%
\left[ 1,6,0\right] }+\frac{30}{r^{2}}D_{\left[ 3,4,0\right] }%
\end{array}%
$

------------------------------------------------------------------------------------------------------

\begin{equation*}
N=8
\end{equation*}

\bigskip $%
\begin{array}{c}
\left( \partial +\frac{2}{r}\right) D_{\left[ 9,0,0\right] }-\frac{16}{r}D_{%
\left[ 7,2,0\right] } \\ 
\left( \partial +\frac{4}{r}\right) D_{\left[ 7,2,0\right] }-\frac{8}{r}D_{%
\left[ 5,4,0\right] } \\ 
\left( \partial +\frac{6}{r}\right) D_{\left[ 5,4,0\right] }-\frac{24}{5r}D_{%
\left[ 3,6,0\right] } \\ 
\left( \partial +\frac{8}{r}\right) D_{\left[ 3,6,0\right] }-\frac{16}{7r}D_{%
\left[ 1,8,0\right] } \\ 
\left( \partial +\frac{10}{r}\right) D_{\left[ 1,8,0\right] }%
\end{array}%
$ $\ 
\begin{array}{c}
\left( \partial _{r}^{2}+\frac{2}{r}\partial _{r}-\frac{16}{r^{2}}\right) D_{%
\left[ 8,0,0\right] }+\frac{112}{r^{2}}D_{\left[ 6,2,0\right] } \\ 
\left( \partial _{r}^{2}+\frac{2}{r}\partial _{r}-\frac{38}{r^{2}}\right) D_{%
\left[ 6,2,0\right] }+\frac{2}{r^{2}}D_{\left[ 8,0,0\right] }+\frac{40}{r^{2}%
}D_{\left[ 4,4,0\right] } \\ 
\left( \partial _{r}^{2}+\frac{2}{r}\partial _{r}-\frac{44}{r^{2}}\right) D_{%
\left[ 4,4,0\right] }+\frac{12}{r^{2}}D_{\left[ 6,2,0\right] }+\frac{72}{%
5r^{2}}D_{\left[ 2,6,0\right] } \\ 
\left( \partial _{r}^{2}+\frac{2}{r}\partial _{r}-\frac{34}{r^{2}}\right) D_{%
\left[ 2,6,0\right] }+\frac{30}{r^{2}}D_{\left[ 4,4,0\right] }+\frac{16}{%
7r^{2}}D_{\left[ 0,8,0\right] } \\ 
\left( \partial _{r}^{2}+\frac{2}{r}\partial _{r}-\frac{8}{r^{2}}\right) D_{%
\left[ 0,8,0\right] }+\frac{56}{r^{2}}D_{\left[ 2,6,0\right] }%
\end{array}%
$

------------------------------------------------------------------------------------------------------

Table 1: \ The isotropic formulas for $\nabla _{\mathbf{r}}\bullet \mathbf{D}%
_{\left[ N+1\right] }$\ are on the left and those for $\nabla _{\mathbf{r}%
}^{2}\mathbf{D}_{\left[ N\right] }$\ are on the right. \ Those derivatives
apply to structure function equations of order $N=4$ to $8$. \ The case $N=8$%
\ was not given by Hill (2001).

\section{\textbf{Comparison with previous results}}

\qquad The expression $\left( \partial _{r}+\frac{2}{r}\right) D_{111}-\frac{%
4}{r}D_{122}$\ in (\ref{2ndorder}) is the same as equation (9) of Yakhot
(2001), and (41) of Hill (1997a). \ The expression $\left( \partial _{r}+%
\frac{2}{r}\right) D_{1111}-\frac{6}{r}D_{1122}$\ in (\ref{HillBor1}) is the
same as in the equation that follows Yakhot's eq.11 , and in equation (16)
of Hill \& Boratav (2001) and in equation (8) of Kurien (2001); $\left(
\partial _{r}+\frac{4}{r}\right) D_{1122}-\frac{4}{3r}D_{2222}$ in (\ref%
{HillBor2}) is the same as in equation (13) of Hill \& Boratav (2001) and
equation (10) of Kurien (2001). \ The expressions $\left( \partial _{r}+%
\frac{2}{r}\right) D_{\left[ 6,0,0\right] }-\frac{10}{r}D_{\left[ 4,2,0%
\right] }$\ and $\left( \partial _{r}+\frac{6}{r}\right) D_{\left[ 2,4,0%
\right] }-\frac{6}{5r}D_{\left[ 0,6,0\right] }$\ for the case $N=5$ in Table
1 are the same as in equations (9) and (10) of Kurien (2001). \ More
generally, the isotropic formulas for $\nabla _{\mathbf{r}}\bullet \mathbf{D}%
_{\left[ N+1\right] }$\ for the case $N_{1}=N$, $N_{2}=N_{3}=0$ are $\left(
\partial _{r}+\frac{2}{r}\right) D_{\left[ N,0,0\right] }-\frac{2\left(
N-1\right) }{r}D_{\left[ N-2,2,0\right] }$\ which agrees with the left-hand
side of equation (7) of Yakhot (2001). \ The other components of $\nabla _{%
\mathbf{r}}\bullet \mathbf{D}_{\left[ N+1\right] }$\ were not given by
Yakhot (2001). \ The expressions from the Laplacian in (\ref{2ndorder}-\ref%
{2ndorder2}) are the same as in (41-42) of Hill (1997a); and (\ref{HillBor3}-%
\ref{HillBor4}) are the same as (7-8) of Hill \& Boratav (2001). \ All of
the remaining results do not appear to have been given previously. \ The
above comparisons are sufficient to verify the matrix algorithm for
generating the structure-function equations to any desired order, as well as
to independently validate the derivation of Yakhot (2001).

\section{\textbf{Summary and discussion}}

\qquad The third paragraph of the introduction summarizes part of this paper
and is not repeated here. \ In addition: \ All of the kinematic
relationships (\ref{kinematic}) between components of isotropic, symmetric
structure functions of arbitrary order have been identified, whereas
previously only one was known. \ All of the components that are zero have
been identified (a recent experimental evaluation of some of them is given
by Kurien \& Sreenivasan 2000). \ The kinematic relationships show that the
scaling exponents of certain different components must be equal; if the
exponents are not equal when evaluated using one's data, then the kinematic
relationships (\ref{kinematic}) provide a measure of either the error in the
exponents or the deviation from local isotropy. \ The dynamic equations of
order $N$ can be used to test the extent of a scaling range for evaluation
of scaling exponents of velocity structure functions of order $N+1$\ because
the time-derivative and viscous terms should be zero in an inertial range. \
The graphical presentations of the balance of Kolmogorov's equation by
Antonia \textit{et al}. (1983), Chambers \& Antonia (1984), Danaila \textit{%
et al}. (1999a,b), and Antonia \textit{et al}. (2000) show the extent of, or
deviation from, inertial-range exponents. \ The higher-order equations given
here can be used in an analogous manner.

\qquad The energy dissipation rate $\varepsilon $\ plays an essential role
at all $r$\ in Kolmogorov's equation. \ In our formulation $\varepsilon $\
arises in (\ref{2ndorder}-\ref{2ndorder2}) from the tensor components $2\nu
E_{11}$\ and $2\nu E_{22}$. \ On the other hand, for the next-order
equations (\ref{HillBor1}-\ref{HillBor2}) Hill (1997b) showed that the
corresponding terms $2\nu E_{111}$\ and\ $2\nu E_{122}$\ are negligible in
the inertial range. \ Yakhot (2001) shows that the components $E_{\left[
N,0,0\right] }$ are negligible in the inertial range for all of the
higher-order equations for which $N$\ is odd. \ Kolmogorov's (1941)
inertial-range scaling using $\varepsilon $ and $r$ as the only relevant
parameters can be used to estimate the relative magnitudes of the term $%
\nabla _{\mathbf{r}}\bullet \mathbf{D}_{\left[ N+1\right] }$ in (\ref{1iso})
to the terms $2\nu \nabla _{\mathbf{r}}^{2}\mathbf{D}_{\left[ N\right] }$
and $2\nu \mathbf{E}_{\left[ N\right] }$. \ Doing so, the ratio of any
nonzero component of $2\nu \nabla _{\mathbf{r}}^{2}\mathbf{D}_{\left[ N%
\right] }$ or $2\nu \mathbf{E}_{\left[ N\right] }$ to the corresponding
component of $\nabla _{\mathbf{r}}\bullet \mathbf{D}_{\left[ N+1\right] }$
is proportional to $\nu /r^{4/3}\varepsilon ^{1/3}=\left( r/\eta \right)
^{-4/3}$, which asymptotically vanishes in the inertial range ($\eta \equiv
\left( \nu ^{3}/\varepsilon \right) ^{1/4}$). \ Thus, both terms $2\nu
\nabla _{\mathbf{r}}^{2}\mathbf{D}_{\left[ N\right] }$ and $2\nu \mathbf{E}_{%
\left[ N\right] }$\ are to be neglected in an inertial range if $N>2$.

\qquad One concludes that all equations of order higher than Kolmogorov's
equation reduce to the isotropic formula for $\nabla _{\mathbf{r}}\bullet 
\mathbf{D}_{\left[ N+1\right] }=-\mathbf{T}_{\left[ N\right] }$ in the
inertial range. \ This formula shows that $\mathbf{T}_{\left[ N\right] }$\
is at the heart of two issues that have received much attention: 1) whether
or not different components of the velocity structure function $\mathbf{D}_{%
\left[ N+1\right] }$\ have differing exponents in the inertial range, and 2)
the increasing deviation of those exponents from Kolmogorov scaling as $N$
increases. \ The physical basis for the importance of $\mathbf{T}_{\left[ N%
\right] }$\ is the importance of vortex tubes to the intermittency
phenomenon (Pullin \& Saffman 1998) combined with the fact that the
pressure-gradient force is essential to the existence of vortex tubes; the
pressure-gradient force prevents a vortex tube from cavitating despite the
centrifugal force. \ Pressure gradients are the sinews of vortex tubes. \
Direct investigation of $\mathbf{T}_{\left[ N\right] }$\ using DNS can
reveal much about the two issues.

\begin{acknowledgement}
The author thanks the organizers of the Hydrodynamics Turbulence Program
held at the Institute for Theoretical Physics, University of California at
Santa Barbara, whereby this research was supported in part by the National
Science Foundation under grant number PHY94-07194. \ The author thanks Prof.
Norbert Peters and Mr. Jonas Boschung who found that the Laplacians on the
right side of Table 1 were in error in my original publication: Hill, R. J.
2001, Equations relating structure functions of all orders. J. Fluid Mech.,
vol. 434,\ pp. 370-388. \ The Laplacians in Table 1 above are correct.
\end{acknowledgement}

\ \ \ \ \ \ \ \ \ \ \ \ \ \ \ \ \ \ \ \ \ \ \ \ \ \ \ \ \ \ \ \ \ \ \ \
REFERENCES

\textsc{Antonia, R. A., Chambers, A. J. \& Browne, L. W. B.} 1983 Relations
between structure functions of velocity and temperature in a turbulent jet. 
\textit{Experiments in Fluids} \textbf{1}, 213-219.

\textsc{Antonia, R. A., Zhou, T. \& Zhu, Y.} 1998 Three-component vorticity
measurements in a turbulent grid flow. \textit{J.~Fluid Mech.} \textbf{374},
29-57.

\textsc{Antonia, R. A., Zhou, T., Danaila, L. \& Anselmet, F.} 2000
Streamwise inhomogeneity of decaying grid turbulence. \textit{Phys. Fluids} 
\textbf{12}, 3086-3089.

\textsc{Arad, I., L}$^{\prime }$\textsc{vov, V. S., \& Procaccia, I.} 1999\
Correlation functions in isotropic and anisotropic turbulence: The role of
the symmetry group. \textit{Phys. Rev. E} \textbf{59}, 6753-6765.

\textsc{Boratav, O. N. \& Pelz, R. B.} 1997 Structures and structure
functions in the inertial range of turbulence. \textit{Phys. Fluids} \textbf{%
9}, 1400-1415.

\textsc{Boratav, O. N.} 1997 On recent intermittency models of turbulence. 
\textit{Phys. Fluids} \textbf{9}, 1206-1208.

\textsc{Camussi, R. \& Benzi, R.} 1997 Hierarchy of transverse structure
functions. \textit{Phys. Fluids} \textbf{9}, 257-259.

\textsc{Chambers, A. J. \& Antonia, R. A.} 1984 Atmospheric estimates of
power-law exponents $\mu $\ and $\mu _{\theta }$. \textit{Bound.-Layer
Meteorol.} \textbf{28}, 343-352.

\textsc{Chen, S., Sreenivasan, K. R., Nelkin, M. \& Cao, N.} 1997 Refined
similarity hypothesis for transverse structure functions in fluid
turbulence. \textit{Phys. Rev. Lett.} \textbf{79}, 2253-2256.

\textsc{Danaila, L., Anselmet, F., Zhou, T. \& Antonia,\ R. A.} 1999a A
generalization of Yaglom's equation which accounts for the large-scale
forcing in heated decaying turbulence. J\textit{. Fluid Mech.} \textbf{391},
359-372.

\textsc{Danaila, L., Dusek, J., Le Gal, P., Anselmet, F., Brun, C. \& Pumir,
A.} 1999b\ Planar isotropy of passive scalar turbulent mixing with a mean
perpendicular gradient. \textit{Phys. Rev. E} \textbf{60}, 1691-1707.

\textsc{Danaila, L., Le Gal, P., Anselmet, F., Plaza, F. \& Pinton, J. F.}
1999c Some new features of the passive scalar mixing in a turbulent flow. 
\textit{Phys. Fluids} \textbf{11}, 636-646.

\textsc{Dhruva, B., Tsuji, Y. \& Sreenivasan, K. R.} 1997 Transverse
structure functions in high-Reynolds-number turbulence. \textit{Phys. Rev. E}
\textbf{56}, R4928-R4930.

\textsc{Grossmann, S., Lohse, D. \& Reeh, A.} 1997 Different intermittency
for longitudinal and transversal turbulent fluctuations. \textit{Phys. Fluids%
} \textbf{9}, 3817-3825.

\textsc{Hill, R. J.} 1997a Applicability of Kolmogorov's and Monin's
equations of turbulence. \textit{J.~Fluid Mech.} \textbf{353}, 67-81.

\textsc{Hill, R. J.} 1997b Pressure-gradient, velocity-velocity structure
function for locally isotropic turbulence in incompressible fluid. NOAA
Tech. Memo. ERL ETL-277, (available from N.T.I.S., 5285 Port Royal Rd.,
Springfield, VA 22161).

Hill, R. J. 2001 Equations relating structure functions of all orders. 
\textit{J. Fluid Mech.}, \textbf{434},\ 370-388.

\textsc{Hill, R. J. \& Boratav, O. N.} 2001 Next-order structure-function
equations. \textit{Phys. Fluids} \textbf{13}, 276-283.

\textsc{Kahaleras, H. Malecot, Y. \& Gagne, Y.} 1996 Transverse velocity
structure functions in developed turbulence. \textit{Advances in Turbulence} 
\textbf{VI}, 235--238.

\textsc{Kerr, R. M., Meneguzzi, M. \& Gotoh, T.} 2001 An inertial range
length scale in structure functions. (submitted \textit{Phys. Fluids},
preprint xxx.lanl.gov/physics/0005004).

\textsc{Kolmogorov, A. N.} 1941 Dissipation of energy in locally isotropic
turbulence. \textit{Dokl. Akad. Nauk SSSR} \textbf{32}, 16.

\textsc{Kurien, S. \& Sreenivasan, K. R.} 2000 Anisotropic scaling
contributions to high-order structure functions in high-Reynolds-number
turbulence. \textit{Phys. Rev. E} \textbf{62}, 2206-2212.

\textsc{Kurien, S. \& Sreenivasan, K. R.} 2001 Mean field approximation in
turbulence theory - Experimental results (preprint).

\textsc{Lindborg, E.} 1996 A note on Kolmogorov's third-order
structure-function law, the local isotropy hypothesis and the
pressure-velocity correlation. \textit{J.~Fluid Mech.} \textbf{326}, 343-356.

\textsc{Lindborg, E.} 1999 Correction to the four-fifths law due to
variations of the dissipation. \textit{Phys. Fluids} \textbf{11}, 510-512.

\textsc{Millionshtchikov, M. D.} 1941 On the theory of homogeneous isotropic
turbulence. \textit{Dokl. Akad. Nauk SSSR} \textbf{32}, 611-614.

\textsc{Monin, A. S. \& Yaglom, A. M.} 1975 \textit{Statistical Fluid
Mechanics: Mechanics of Turbulence}, vol. 2. the MIT Press.

\textsc{Nelkin, M.} 1999 Enstrophy and dissipation must have the same
scaling exponent in high Reynolds number turbulence. \textit{Phys. Fluids} 
\textbf{11}, 2202-2204.

\textsc{Noullez, A., Wallace, G., Lempert, W., Miles, R. B. \& Frisch, U.}
1997 Transverse velocity increments in turbulent flow using the relief
technique. \textit{J.~Fluid Mech.} \textbf{339}, 287-307.

\textsc{Pullin, D. I. \& Saffman, P. G.} 1998 Vortex dynamics in turbulence. 
\textit{Annu. Rev. Fluid Mech.} \textbf{30}, 31-51.

\textsc{Pumir, A. \& Shraiman, B. I.} 1995 Persistent small scale anisotropy
in homogeneous shear flows. \textit{Phys. Rev. Lett.} \textbf{75}, 3114-3117.

\textsc{Shen, X. \& Warhaft, Z.} 2000 The anisotropy of the small scale
structure in high Reynolds number ($R_{\lambda }\sim 1000$) turbulent shear
flow. \textit{Phys. Fluids} \textbf{12}, 2976-2989.

\textsc{Su, L. K. \& Dahm, W. J. A.} 1996 Scalar imaging velocimetry
measurements of the velocity gradient tensor field in turbulent flows. I.
Experimental results. \textit{Phys. Fluids} \textbf{8}, 1883-1906.

\textsc{van de Water, W. \& Herweijer, J. A.} 1999 Higher-order structure
functions of turbulence. \textit{J.~Fluid Mech.} \textbf{387}, 3-37.

\textsc{Yakhot, V.} Mean-field approximation and a small parameter in
turbulence theory. (accepted, \textit{Phys. Rev. E}, 2001).

\textsc{Zhou, T. \& Antonia, R. A.}\ 2000\ Reynolds number dependence of the
small-scale structure of grid turbulence. \textit{J.~Fluid Mech.} \textbf{406%
}, 81-107.

\begin{center}
\textit{the end}
\end{center}

\end{document}